\pdfoutput=1
\documentclass[11pt,twoside,a4paper,cmspaper,final,collab]{cms-tdr}
\def\svnVersion{41f8024}\def\svnDate{2023/07/13}\def\cmsCernNoTag{CERN-EP-2022-037}\def\cmsCernDate{\today}\def\cmsMessage{Published in Physical Review D as \href{http://dx.doi.org/10.1103/PhysRevD.108.012009}{\doi{10.1103/PhysRevD.108.012009}.}}
\begin{document}\cmsNoteHeader{EXO-20-008}

\ifthenelse{\boolean{cms@external}}{\providecommand{\cmsLeft}{top\xspace}}{\providecommand{\cmsLeft}{left\xspace}}
\ifthenelse{\boolean{cms@external}}{\providecommand{\cmsRight}{bottom\xspace}}{\providecommand{\cmsRight}{right\xspace}}

\newcommand{\Bstar}{\ensuremath{\PQb^{\ast}}\xspace}
\newcommand{\Qstar}{\ensuremath{\PQq^{\ast}}\xspace}
\newcommand{\mjj}{\ensuremath{m_{\mathrm{jj}}}\xspace}
\newcommand{\gH}{\ensuremath{g_\text{H}}\xspace}
\newcommand{\gF}{\ensuremath{g_\text{F}}\xspace}
\newcommand{\gV}{\ensuremath{g_\text{V}}\xspace}
\newcommand{\cH}{\ensuremath{c_\text{H}}\xspace}
\newcommand{\cF}{\ensuremath{c_\text{F}}\xspace}

\cmsNoteHeader{EXO-20-008}

\title{Search for narrow resonances in the \texorpdfstring{\PQb}{b}-tagged dijet mass spectrum in proton-proton collisions at \texorpdfstring{$\sqrt{s} = 13\TeV$}{sqrt(s) = 13 TeV}}

\date{\today}

\abstract{A search is performed for narrow resonances decaying to final states of two jets, with at least one jet originating from a \PQb quark, in proton-proton collisions at $\sqrt{s} = 13\TeV$. The data set corresponds to an integrated luminosity of 138\fbinv collected with the CMS detector at the LHC. Jets originating from energetic \PQb hadrons are identified through a \PQb-tagging algorithm that utilizes a deep neural network or the presence of a muon inside a jet. The invariant mass spectrum of jet pairs is well described by a smooth parametrization and no evidence for the production of new particles is observed. Upper limits on the production cross section are set for excited \PQb quarks and other resonances decaying to dijet final states containing \PQb quarks. These limits exclude at 95\% confidence level models of \PZpr bosons with masses from 1.8 to 2.4\TeV and of excited \PQb quarks with masses from 1.8 to 4.0\TeV. This is the most stringent exclusion of excited \PQb quarks to date.}

\hypersetup{
pdfauthor={CMS Collaboration},%
pdftitle={Search for narrow resonances in the b-tagged dijet mass spectrum in proton-proton collisions at  sqrt(s)=13 TeV},%
pdfsubject={CMS},%
pdfkeywords={CMS,  EXO, Z', excited, bottom, dijet}%
}

\maketitle

\section{Introduction}

Searches for heavy resonances decaying into jet pairs provide a powerful tool with the potential for discovering new physics at hadron colliders. These resonances can be produced via parton-parton interactions, and then can decay to two partons. The final-state partons hadronize and are observed as two jets, referred to as dijets.

Such heavy resonances at the \TeV scale are a feature of several models that extend the standard model (SM) to address some of its shortcomings, such as the extreme fine tuning required in quantum corrections to accommodate a Higgs boson observed at a mass of 125\GeV~\cite{Khachatryan:2014jba,Aad:2014aba,Aad:2013xqa,Aad:2015zhl}.
These models introduce heavy resonances that couple to the SM bosons and fermions~\cite{Barger:1980ix,1126-6708-2009-11-068}. A minimal extension of the SM is represented by the sequential standard model (SSM)~\cite{Langacker_2009}, which introduces a spin-1 \PZpr boson with the same couplings to fermions as the SM \PZ boson, but with a much larger mass. The SSM, among many others, is generalized in the heavy vector triplet (HVT) framework~\cite{Pappadopulo2014}, which extends the SM by introducing a triplet of heavy vector bosons, one neutral \PZpr and two oppositely charged \PWpr, collectively referred to as $\PV^\prime$.
The heavy vector bosons couple to SM bosons and fermions with strengths $\gH=\gV\cH$ and $\gF=g^2\cF/\gV$, respectively, where \gV is the strength of the new interaction; \cH is the coupling between an HVT boson, the Higgs boson, and a longitudinally polarized SM vector boson; \cF is the coupling between an HVT boson and an SM fermion; and $g$ is the electroweak coupling.
In this search, we consider two benchmarks, defined in Ref.~\cite{Pappadopulo2014}. In the Model~A scenario of the HVT framework, the coupling strengths of the heavy vector bosons to SM bosons and fermions are of the same order, and the new particles decay primarily to fermions. In the Model~B scenario of the same framework, the couplings to fermions are suppressed with respect to the couplings to bosons, resulting in a branching fraction of the new particles to SM bosons that is close to unity.

Other models extending the SM postulate that leptons and quarks are composite objects, composed of more fundamental constituents. At an energy beyond the scale of constituent binding energies, the compositeness scale $\Lambda$, a new interaction should emerge~\cite{TERAZAWA1982387,Eichten1983,HARARI1984159}. Such models foresee the existence of excited states of quarks (\Qstar) and leptons ($\ell^{\ast}$), which could be produced in high-energy collisions and then detected through their decays to SM particles. Excited states of composite quarks~\cite{Baur:1987ga}, which would be produced through a gauge interaction or via contact interactions (CI)~\cite{Baur:1989kv}, can result in large cross sections and could decay predominantly to a quark and a gluon ($\PQq\Pg$). In the model considered in this search, the compositeness scale $\Lambda$ is set equal to the resonance mass and the couplings of excited quarks to other particles are assumed to be the same as for nonexcited fermions.

Since the early 1980s, searches for resonances in the dijet invariant mass spectrum~\cite{Harris:2011bh} have been common at hadron colliders.
From proton-proton~($\Pp\Pp$) collision data at $\sqrt{s}=7$ and 8\TeV, the CMS Collaboration has published searches for dijet resonances where at least one of the two jets in the final state arises from a \PQb quark~\cite{Chatrchyan_2013,PhysRevD.91.052009,PhysRevLett.120.201801}, based on a jet flavor identification commonly referred to as \PQb tagging.
The most recent search from the CMS Collaboration at $\sqrt{s}=13\TeV$~\cite{Sirunyan:2019vgj} did not attempt to identify the type of final-state parton that produced the jets, and considered resonance models where the final-state partons could be gluons or any flavor of SM quarks.
Similar to a recent search from the ATLAS experiment~\cite{Aad:2019hjw}, the present search identifies heavy-flavored quarks among the two final-state partons, and has increased sensitivity to models of dijet resonances that decay to \PQb quarks.
Examples are such models that predict an excited \PQb quark (\Bstar), or a \PZpr where the couplings to the third generation are enhanced relative to the couplings to the first and second generations. Such enhanced couplings are also favored in models created to accommodate the possible anomalies observed in the low-energy heavy-flavor sector of the SM~\cite{Ciezarek:2017yzh,LHCb:2021trn}.

In this paper we report a search for \PZpr decaying into a \PQb quark pair as well as \Bstar decaying into \PQb quark and gluon, where two production modes are considered for the latter; $\PQb\Pg\to\Bstar$ and \Bstar via CI. Both resonances are assumed to be narrow. The data used for the analysis were collected by the CMS detector from $\Pp\Pp$ collisions at a center-of-mass energy of 13\TeV during the LHC Run 2 (2016--2018), and correspond to a total integrated luminosity of 138\fbinv.

Tabulated results are provided in the HEPData record for this analysis~\cite{HEPData}.

\section{The CMS detector}\label{sec:detector}

The central feature of the CMS apparatus is a superconducting solenoid of 6\unit{m} internal diameter, providing a magnetic field of 3.8\unit{T}. Within the solenoid volume are a silicon pixel and strip tracker, a lead tungstate crystal electromagnetic calorimeter (ECAL), and a brass and scintillator hadron calorimeter (HCAL), each composed of a barrel and two endcap sections. The ECAL and HCAL provide pseudorapidity coverage up to $\abs{\eta}<3.0$, which is further extended by forward calorimeters~\cite{Chatrchyan:2008zzk}. Muons are measured in gas-ionization detectors embedded in the steel flux-return yoke outside the solenoid. For nonisolated particles of transverse momentum $1<\pt<10\GeV$ and $\abs{\eta}<1.4$, the reconstructed tracks have a \pt resolution of 1.5\%, a transverse impact parameter resolution of 25--90\mum, and a longitudinal impact parameter resolution of 45--190\mum~\cite{Chatrchyan:2014fea}.

Events of interest are selected using a two-tiered trigger system. The first level, composed of custom hardware processors, uses information from the calorimeters and muon detectors to select events at a rate of around 100\unit{kHz} within a fixed latency of about 4\mus~\cite{Sirunyan:2020zal}. The second level, known as the high-level trigger, consists of a farm of processors running a version of the full event reconstruction software optimized for fast processing, and reduces the event rate to around 1\unit{kHz} before data storage~\cite{Khachatryan:2016bia}.

A detailed description of the CMS detector, together with a definition of the coordinate system used and the relevant kinematic variables, can be found in Ref.~\cite{Chatrchyan:2008zzk}.

\section{Data and simulated samples}\label{sec:mcsimulation}

The data sample analyzed has been collected by the CMS experiment with trigger algorithms that require large hadronic activity in the event. In particular, two sets of trigger algorithms are used: single-jet triggers that require a jet with $\pt>550\GeV$ to be present in the event and triggers requiring the scalar \pt sum for all jets with $\pt>30\GeV$ and $\abs{\eta}<3.0$ in the event, to exceed 1050\GeV.

The $\qqbar\to\PZpr\to\PQb\PAQb$ signal Monte Carlo (MC) simulated events are produced at leading-order (LO) accuracy in quantum chromodynamics (QCD) with the \MGvATNLO~v2.4.2 matrix element generator~\cite{MadGraph}. Various \PZpr mass hypotheses in the range of 1.6 to 8\TeV are considered for these narrow resonance signals, which have a natural width smaller than the experimental dijet mass resolution (2--3\%). This approximation is valid in a large fraction of the HVT parameter space, and is fulfilled in the Model~A and Model~B benchmarks discussed earlier~\cite{Pappadopulo2014}.

\begin{figure}[!htb]\centering
  \includegraphics[width=.49\textwidth]{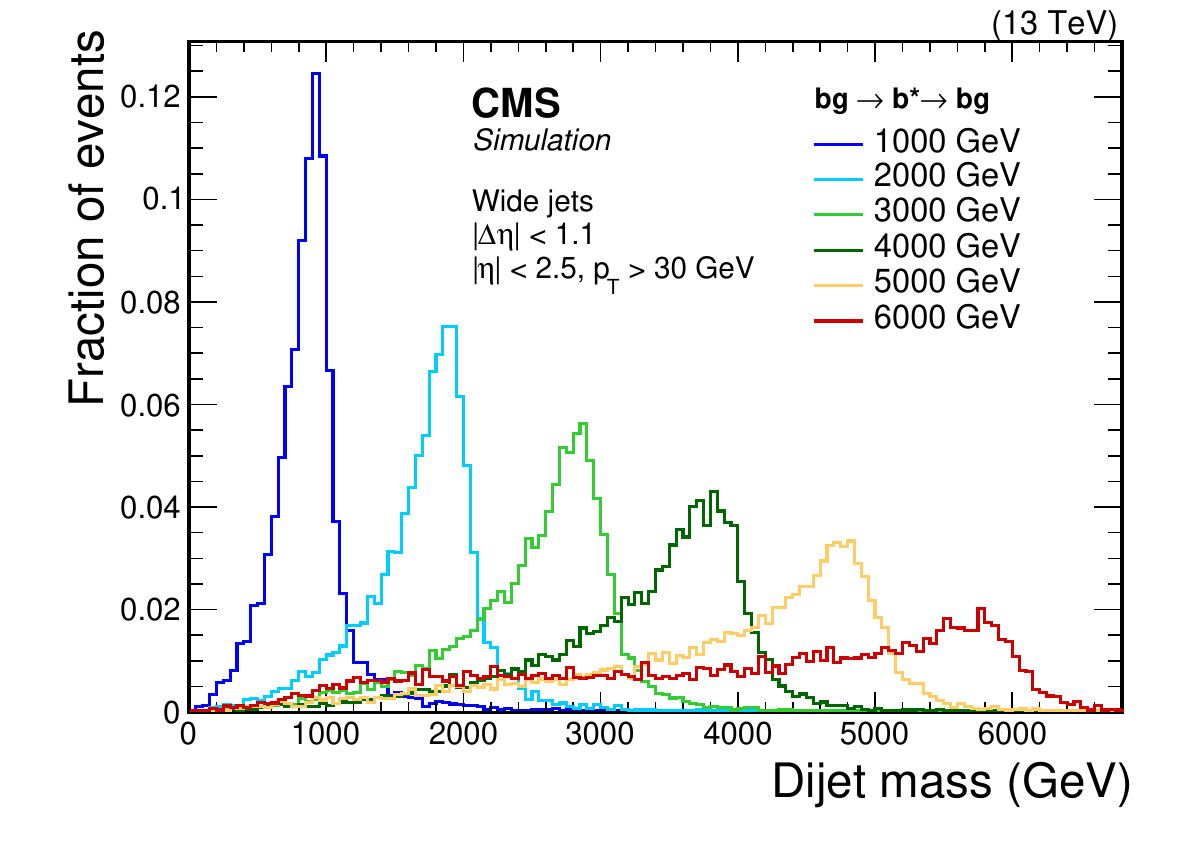}
  \caption{Simulated signal shapes of \Bstar from the process $\PQb\Pg\to\Bstar\to\PQb\Pg$. Shown are the wide jets (see text) used to reconstruct the dijet mass spectra.}
  \label{fig:signal_shape}
\end{figure}

For the \Bstar production processes, the signal samples are produced with \PYTHIA~8.230~\cite{Sjostrand:2014zea} for resonance masses up to 6\TeV. The characteristic dijet mass shape corresponding to the $\PQb\Pg\to\Bstar\to\PQb\Pg$ signal component is reported in Fig.~\ref{fig:signal_shape} for resonance masses from 1 to 6\TeV. The contribution of the low-mass tail to the signal shape becomes more dominant at large resonance masses. This is because the resonance natural mass shape coming from a Breit-Wigner distribution is convolved with falling parton distribution functions (PDFs), where for larger resonance masses the PDFs fall more steeply.

In all cases parton showering and hadronization processes are simulated by interfacing the event generators with~\PYTHIA~8.230 with the CUETP8M1~\cite{Skands:2014pea,Khachatryan:2015pea} tune. The NNPDF2.3LO~\cite{2013290,BALL2012153} PDFs are used to model the momentum distribution of the colliding partons inside the protons. The generated events include additional $\Pp\Pp$ interactions (pileup) with a distribution that is chosen to match that observed in the data.
These events are processed through a full detector simulation based on \GEANTfour~\cite{Agostinelli:2002hh} and reconstructed with the same algorithms as those used for data.

\section{Event reconstruction and selection}\label{sec:reconstruction}

The event reconstruction is performed using a particle-flow (PF) algorithm~\cite{Sirunyan:2017ulk}, which uses an optimized combination of information from the various elements of the CMS detector to reconstruct and identify individual particles produced in each collision. The algorithm identifies each reconstructed particle either as an electron, a muon, a photon, or a charged or neutral hadron.
The charged component of the pileup contribution is removed by applying the charged hadron subtraction mitigation algorithm. The momenta of neutral hadrons are rescaled according to their probability to originate from the primary interaction vertex deduced from the distribution of associated energy deposits in the calorimeters~\cite{Sirunyan:2020foa}. The PF candidates are then clustered into jets using the anti-\kt algorithm~\cite{Cacciari:2008gp} implemented in the \FASTJET package~\cite{Cacciari:2011ma,Cacciari:2008gn} with a distance parameter $R=0.4$.

Jet energy corrections, extracted from simulation and data in multijet, $\gamma+\text{jets}$, and $\PZ+\text{jets}$ events, are applied as a function of \pt and $\eta$ to correct the jet response and to account for residual differences between data and simulation. Multijet events refer to SM events composed uniquely of jets produced through the strong interaction. The energy resolution for jets with $\pt\sim1\TeV$ is approximately 5\%~\cite{Khachatryan:2016kdb}. Jets are required to pass identification criteria, which have negligible impact on the signal efficiency, in order to remove spurious jets arising from detector noise~\cite{CMS-PAS-JME-16-003}.

Events are required to contain at least two jets, each with $\pt>30\GeV$ and $\abs{\eta}<2.5$, and each separated by $\Delta R=\sqrt{(\Delta\eta)^{2}+(\Delta\phi)^{2}}>0.4$ from any electrons or muons. Here, $\phi$ is the azimuthal angle in radians. Geometrically close jets are combined into ``wide jets'', and used to determine the dijet mass, as in our previous dijet searches~\cite{Sirunyan:2019vgj}. In forming wide jets, the two jets with the largest \pt are used as seeds, and the four-momenta of jets within $\Delta R<1.1$ of a seed jet are added to the seed jet's four-momentum. The wide-jet algorithm, designed for dijet resonance event reconstruction, reduces the analysis sensitivity to gluon radiation from the final-state partons.

The pseudorapidity separation of the two wide jets is required to be $\abs{\Delta\eta}<1.1$. This requirement suppresses $t$-channel multijet production, and enhances the contribution to the signal from $s$-channel production.

In order to ensure the full efficiency of the trigger, only events with a dijet invariant mass \mjj larger than 1530\GeV are considered. The trigger efficiency measurement is performed with a sample acquired with an independent trigger algorithm that requires the presence of a muon with $\pt>50\GeV$ at the high-level trigger.

Jets from \PQb quarks are identified using a \PQb-tagging DeepJet discriminator that relies on the application of convolutional neural networks on low-level objects, such as PF candidates and secondary vertex information, and improves the jet flavor identification capabilities of the CMS experiment, especially at high jet momenta~\cite{Bols:2020bkb,CMS-DP-2018-058}.
The operation point of the discriminator was chosen in order to maximize the sensitivity to the \PZpr and \Bstar signal searches. It corresponds to a misidentification probability of 1\% for jets from light (\PQu, \PQd, \PQs) quarks and gluons with \pt of about 80\GeV. For this mistagging value, the efficiency of identifying genuine \PQb quarks is approximately 75, 45, and 10\% for a jet with \pt of 0.2, 1.0, and 3.0\TeV, respectively. The simulated event distributions passing the tagger are adjusted via scale factors such that they agree with the observed data.

\begin{figure}[!htb]\centering
  \includegraphics[width=.49\textwidth]{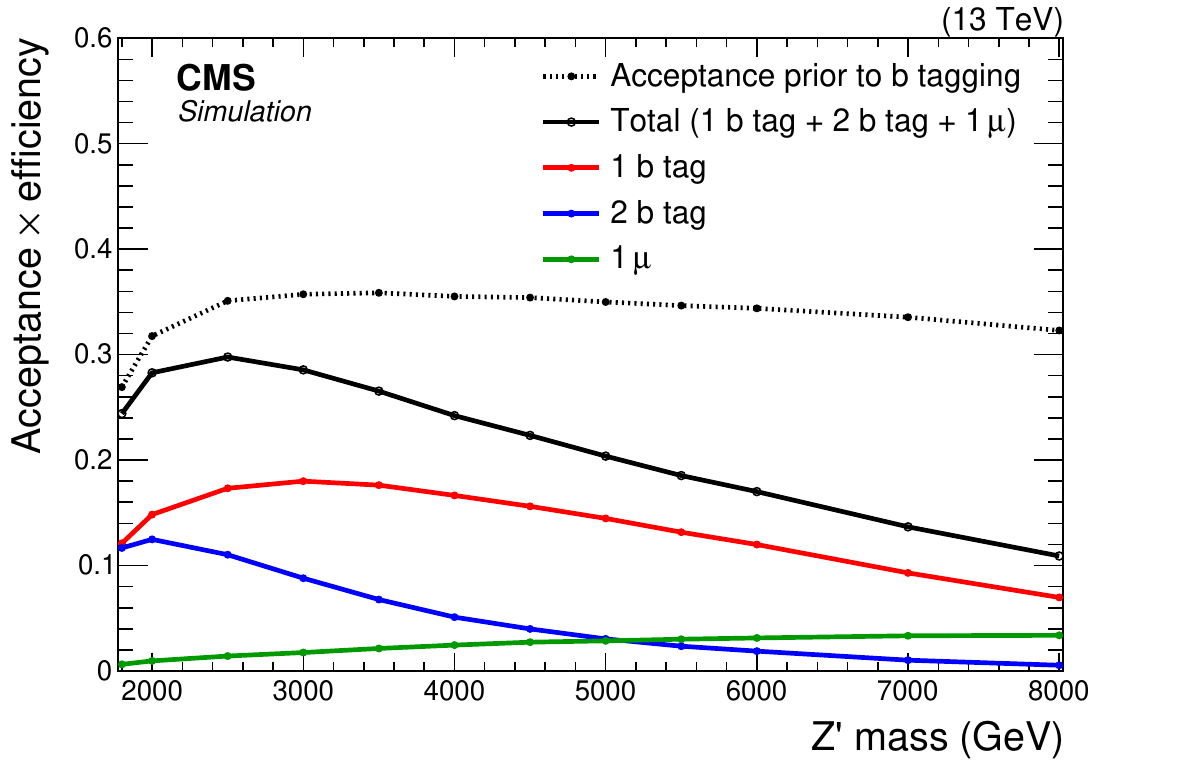}
  \caption{The product of acceptance and efficiency of the event selection for a $\PZpr\to\bbbar$ resonance as a function of the resonance mass.}
  \label{fig:efficiency}
\end{figure}

Events are divided into independent categories, depending on the signal model considered. When searching for a \Bstar resonance decaying to a \PQb quark and a gluon, only a single category with at least one \PQb-tagged jet is needed. For a \PZpr signal, three categories are defined: 2\PQb, 1\PQb and muon. Events are included in the 2\PQb category if both jets fulfill the \PQb-tagging requirements, and in the 1\PQb category if only one of the two jets is \PQb tagged. The muon category contains the events in which neither jet passes the \PQb tag selection, but at least one jet contains a muon, identified within both the inner tracker and the muon detector. The presence of the muon naturally enriches the heavy-flavor component of the jets, mitigating the loss of signal efficiency of the \PQb-tagging algorithm at very large \pt. This category dominates the signal sensitivity for \PZpr mass larger than 5\TeV, a region where the muon presence request has a roughly 50\% larger signal efficiency than in the 2\PQb category. The product of acceptance and efficiency of the selections for the signal were very similar in all three data-taking years, and the average values are shown in Fig.~\ref{fig:efficiency}. The dominant source of inefficiency is the requirement $\abs{\Delta\eta}<1.1$, which reduces the acceptance to approximately 41\%.

\section{Background estimation}\label{sec:bkg}

Background estimates based on simulated data predict that it is largely dominated by multijet production, which accounts for more than 95\% of the total background. The contribution of top quark pair production is approximately 3--4\% of the total background, depending on the \PQb-tagging category. Production of vector bosons in association with partons, and production of boson pairs, contribute the remaining 1--2\% of the total background.

\begin{figure*}[!htbp]\centering
  \includegraphics[width=\textwidth]{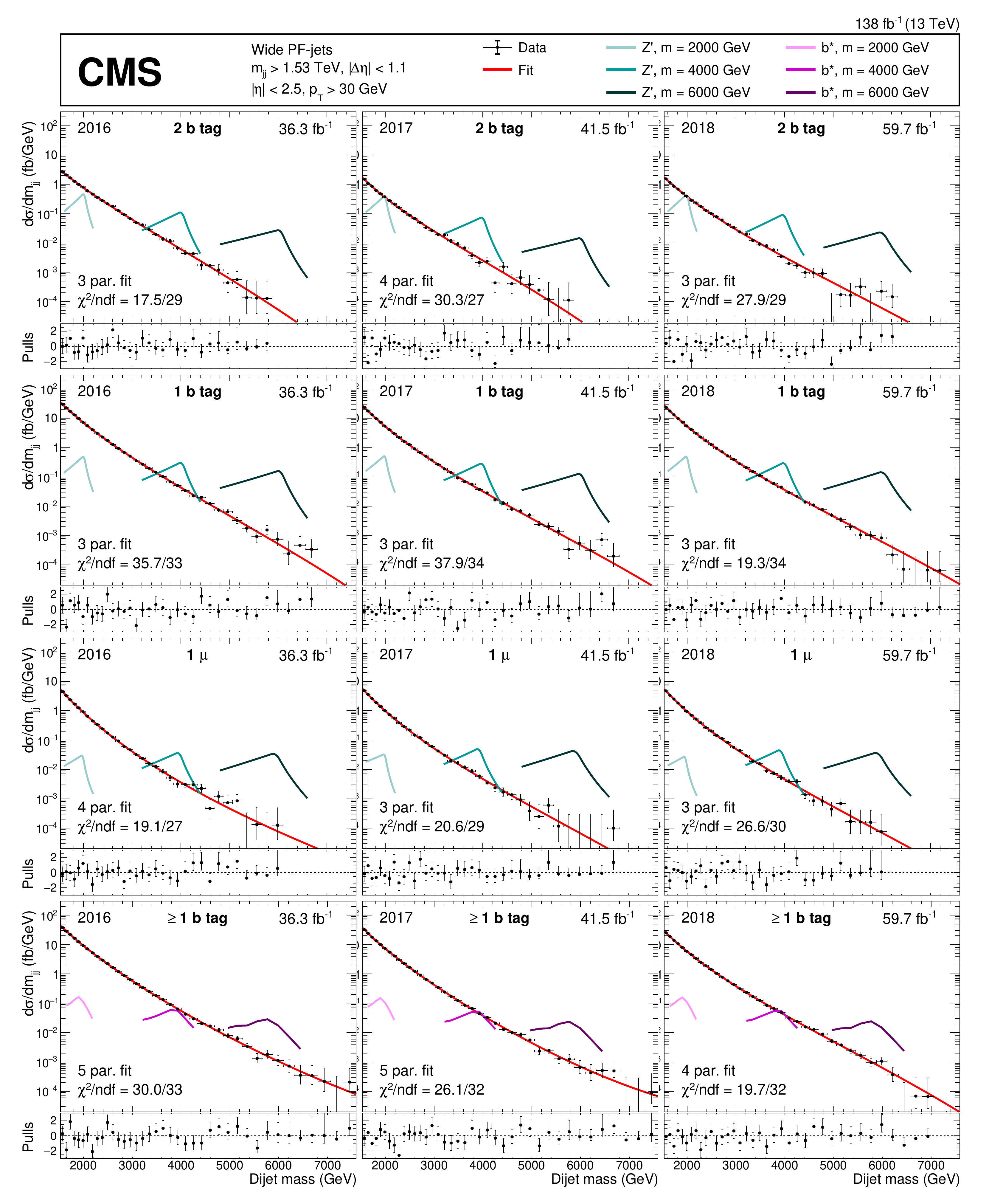}
  \caption{The observed differential cross sections as a function of the dijet mass, shown as fit with the background functions, for the four tagging categories (rows) and the three data-taking periods (columns). The number of parameters in the fit, and the goodness of fit ``$\chi^2/\text{ndf}$'', are listed where ``ndf'' is the number of degrees of freedom. The lower panel within each row shows the pulls, $(\text{data}-\text{fit})/\text{uncertainty}$, in units of the statistical uncertainty in data. The upper three rows are used to search for \PZpr models, the bottom row is used to search for the \Bstar model, and example shapes of these signal models are shown with the same arbitrary normalization for three choices of resonance mass.}\label{fig:bkg_fits}
\end{figure*}

The background is estimated directly from data assuming that it can be described by a smooth, monotonically decreasing function. The validity of this assumption is checked in simulation.
The functions considered are power laws of the variable $x=\mjj/\sqrt{s}$, where $\sqrt{s}=13\TeV$.
Starting from the simplest functional form, an iterative procedure based on the Fisher F-test~\cite{Ftest} is used to check at 90\% confidence level (\CL) if additional parameters are needed to model the individual background distributions.
Depending on the dataset considered for the fit, a three-parameter ($p_{0}(1-x)^{p_{1}}/x^{p_{2}}$),  four-parameter ($p_{0}(1-x)^{p_{1}}/x^{p_{2}+p_{3}\log(x)}$) or five-parameter ($p_{0}(1-x)^{p_{1}}/x^{p_{2}+p_{3}\log(x)+p_{4}\log^{2}(x)}$) functional form is necessary to describe the data.
The observed dijet mass spectra are well modeled by the background functional forms, as shown in Fig.~\ref{fig:bkg_fits}, where the widths of the dijet mass bins correspond to the dijet mass resolution.

The expected shape of the reconstructed signal mass distribution is extracted from the simulated signal samples. The \Bstar signal shapes are dijet mass distributions obtained directly from the simulation, as shown in Fig.~\ref{fig:signal_shape}, while the \PZpr signal shapes are parametrizations that are fit separately to the simulation samples for each category, using a double-sided Crystal Ball function~\cite{CrystalBallRef}. This functional form consists of a Gaussian distribution that models the core of the shape, and two power law functions that model the upper and lower tails, and requires a total of six parameters. The resolution of the reconstructed \PZpr is given by the width of the Gaussian core, and is found to be constant at 2\% of the resonance mass. The signal shape, for a \PZpr of arbitrary mass, is then obtained by a spline interpolation of the simulated signal shape parameter values as a function of mass. Examples of the \PZpr and \Bstar shapes are shown in Fig.~\ref{fig:bkg_fits}, with arbitrary normalizations.

Statistical tests have been performed to check the robustness of the fit method. Pseudoexperiments are generated after injecting a simulated signal with a range of values for the signal mass and cross section. The dijet mass distribution of the pseudodata is then fitted with the signal distribution combined with an alternate background function containing one more parameter than the nominal function. The fitted signal yield is found to be compatible within one third of the statistical uncertainty to the injected yield, regardless of the injected signal strength and resonance mass. These tests confirm that the background estimation is insensitive both to the choice of the function used and to the possible presence of a signal in the data.

\section{Systematic uncertainties}
\label{sec:sys}

The background shape is estimated from the fit to the data in the considered categories, and no assumption or constraint is applied to function parameters. These parameters are considered as uncorrelated among categories and data-taking periods. As described in Sec.~\ref{sec:bkg}, different parametrizations of the fit function were studied and introduce no additional source of uncertainty.

The dominant uncertainties in the signal arise from the \PQb tagging and the jet-reconstruction uncertainties. The uncertainty in the \PQb-tagging scale factors~\cite{CMS:2017wtu} yields an uncertainty in the signal normalization of 2 and 4\% in the 1\PQb and ${\geq}1\PQb$ categories, respectively, and 15\% in the 2\PQb and muon categories.
Uncertainties in the reconstruction of the hadronic jets affect mainly the shape of the reconstructed resonance mass. The four-momenta of the reconstructed jets are scaled and smeared according to the uncertainties in the jet \pt and momentum resolution. These effects result in a 2\% uncertainty in the mean, and 10\% in the width of the signal.
The signal normalization is also affected by the uncertainty in the selection of the muons inside the jets. The efficiency of the identification of the muons inside jets is measured in a statistically independent and almost pure sample of high-\pt \PQb~jets, originating from the decay of pair-produced top quarks. This uncertainty is deduced from the statistical uncertainty in the efficiency measurement, ranging up to 43\% for muons with $\pt<100\GeV$. An uncertainty of 100\% is assumed for jets with the \pt of an associated muon beyond this 100\GeV threshold, because no data was available in this region.
Additional systematic uncertainties affecting the signal normalization include the vetoes for high-\pt leptons and missing transverse momenta (accounting for 1\% each), pileup contributions (0.1\%), and the integrated luminosity (1.2\% in 2016~\cite{CMS:2021xjt}, 2.3\% in 2017~\cite{CMS:2018elu}, and 2.5\% in 2018~\cite{CMS:2019jhq}). The systematic uncertainty from the choice of PDFs~\cite{Butterworth:2015oua} is estimated to be 8--41\% of the normalization of the signal cross section, depending on the resonance mass. The factorization and renormalization scale uncertainties are estimated by varying the scales up and down by a factor of 2, both simultaneously and independently, using the maximum obtained value. The resulting effect is a variation of 6--14\% of the normalization of the signal cross section.

\section{Results and interpretation}
\label{sec:res}

Exclusion limits are obtained by performing a background-only fit and a combined signal-plus-background fit to the dijet mass distributions, separately by category and data-taking year. In the fit, based on a profile likelihood, the parameters and the normalization of the background in each category are left free to float. The systematic uncertainties in the signal are treated as nuisance parameters, with Gaussian constraints for the jet energy scale and resolution, and log-normal constraints for the integrated luminosity, and are profiled in the statistical interpretation~\cite{CMS-NOTE-2011-005}.
The uncertainties that affect the signal normalization (PDFs and factorization and renormalization scales) are treated differently depending on how the exclusion is presented. When deriving upper limits on the cross section, these uncertainties are not varied in the fit, but are reported separately as the uncertainty in the theoretical cross sections from the model. When placing limits on the model parameters, these nuisance parameters are fixed at the best-fit values, in the same manner as with the other systematic uncertainties.
A more detailed description of the statistical treatment and the likelihood function is reported in Ref.~\cite{Sirunyan:2019vgt}. Upper limits at 95\% \CL are set using the \CLs modified frequentist method~\cite{CLS1,CLS2}, adopting the asymptotic approximation~\cite{Asymptotic}.

\begin{figure}[!htb]\centering
  \includegraphics[width=.49\textwidth]{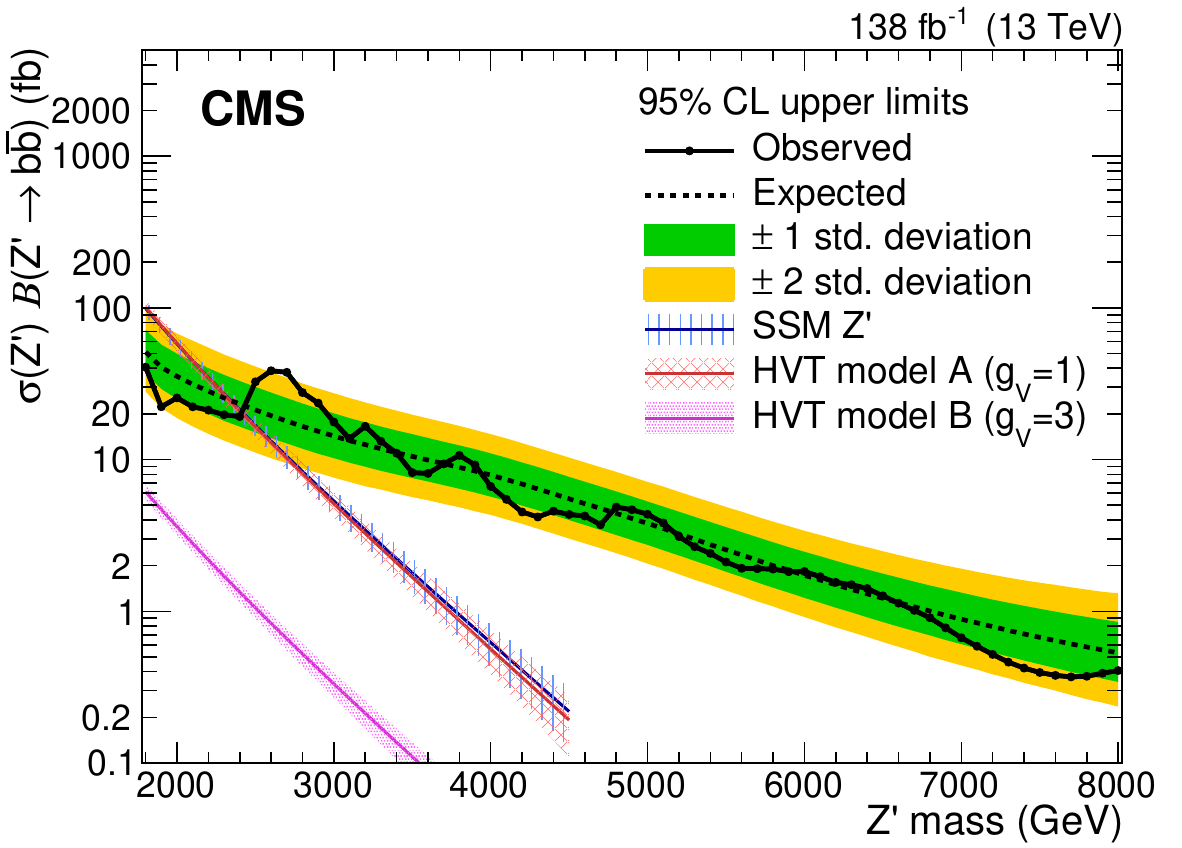}
  \includegraphics[width=.49\textwidth]{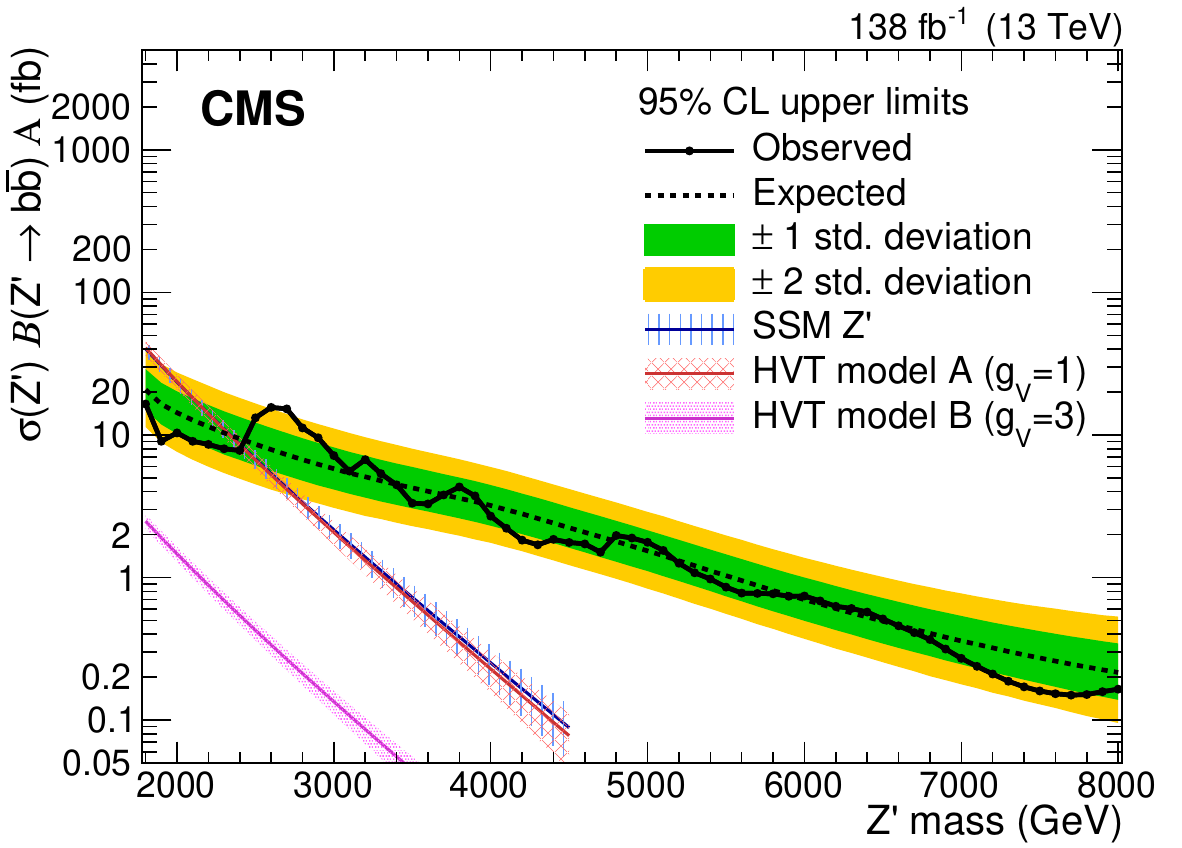}
  \caption{The observed 95\% \CL upper limits (solid curve) on the product of the cross section and branching fraction (\cmsLeft), and multiplied by signal acceptance (\cmsRight), for a resonance decaying to \bbbar. The corresponding expected limits (dashed curve) and their variations at the 1- and 2-standard deviation levels (shaded bands) are also shown. Limits are compared to predicted cross sections for \PZpr bosons from the sequential SM (SSM) and the heavy vector triplet (HVT) models~A and B. The latter two models follow the parameter choices of $\gV=1$ and $\gV=3$ respectively.}
  \label{fig:lim_Zp}
\end{figure}

A model-independent representation of the observed upper limit on the product of cross section and branching fraction of a \bbbar resonance, as well as the expected limit and its relative 68 and 95\% uncertainty bands, are shown in Fig.~\ref{fig:lim_Zp}. The acceptance, which is included in the product on the \cmsRight side of Fig.~\ref{fig:lim_Zp}, is defined exclusively via the geometric requirements $\pt>30\GeV$, $\abs{\eta}<2.5$, and $\abs{\Delta\eta}<1.1$. This analysis sets an observed mass limit of 2.4\TeV, and an expected mass limit of 2.3\TeV, at 95\% \CL, on a narrow \PZpr resonance in both the SSM and the HVT Model~A. No upper limit can be set on Model~B with suppressed fermionic coupling.

\begin{figure}[!htb]\centering
  \includegraphics[width=0.49\textwidth]{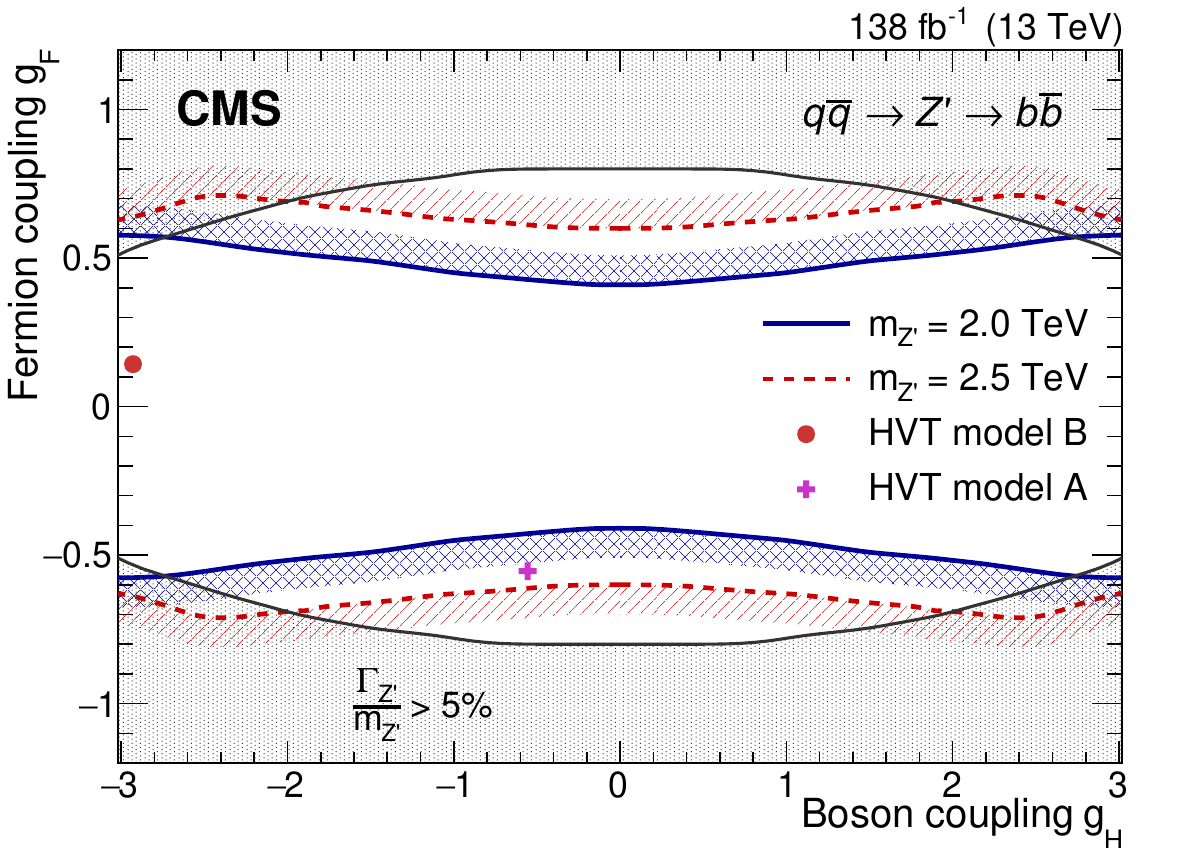}
  \caption{The coupling strengths to SM bosons (\gH) and fermions (\gF) of a \PZpr boson with mass 2.0\TeV (blue) and 2.5\TeV (red) that are excluded at 95\% \CL for the HVT model. The shading indicates the excluded side of the contour. The benchmark scenarios corresponding to HVT models~A and~B are represented by a purple cross and a red point, respectively. The gray shaded area corresponds to the region where the resonance natural width is predicted to be larger than the typical experimental resolution, and thus the narrow-width approximation is not fulfilled.}
  \label{fig:hvt}
\end{figure}

The cross section limit shown in Fig.~\ref{fig:lim_Zp} (\cmsLeft), is reinterpreted in Fig.~\ref{fig:hvt} as limits on the coupling strengths of a heavy vector boson to SM bosons (\gH) and fermions (\gF). The subset of the parameter space where the natural width of the resonance is larger than the typical experimental resolution, and hence the narrow width approximation is invalid, is indicated in Fig.~\ref{fig:hvt} with a gray shaded area.

\begin{figure}[!htb]\centering
  \includegraphics[width=.49\textwidth]{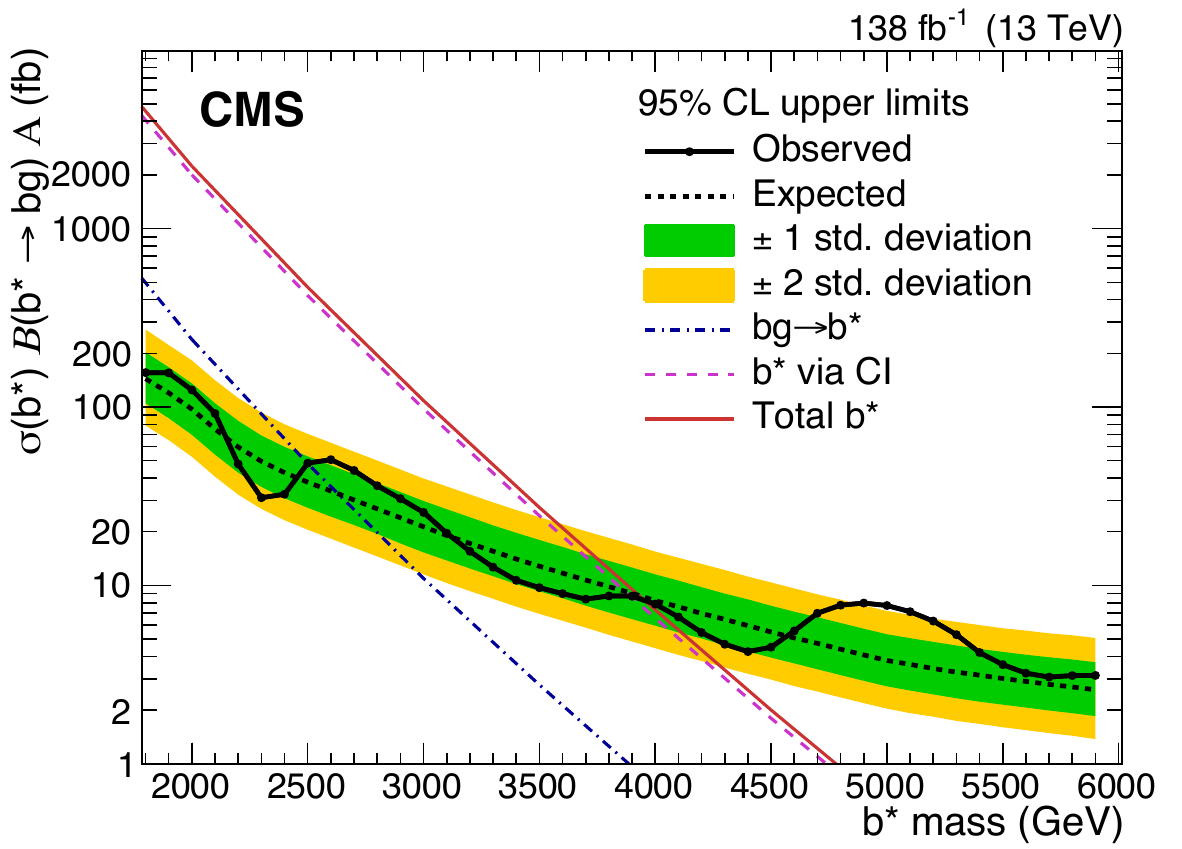}
  \caption{The observed 95\% \CL upper limits on the product of the cross section, branching fraction, and acceptance for dijet resonances decaying to a \PQb quark and a gluon (points). The corresponding expected limits (short dashed) and their variations at the 1- and 2-standard deviation levels (shaded bands) are also shown. Limits are compared to predictions for single \Bstar production (blue, dot dashed), the resonant component of the \Bstar production via contact interactions (magenta, long dashed), and the total \Bstar signal from the sum of these two production modes (red, solid).}
  \label{fig:bStarCombinedLimit}
\end{figure}

For resonances that decay to a \PQb quark and a gluon, upper limits are set by using the category where at least one of the two leading jets is \PQb tagged. Upper limits on the joint product of cross section, branching fraction, and acceptance are reported in Fig.~\ref{fig:bStarCombinedLimit}. When considering the single \Bstar production process, $\PQb\Pg\to\Bstar\to\PQb\Pg$, with the cross section shown in Fig.~\ref{fig:bStarCombinedLimit}, this analysis sets 95\% \CL mass limits of 2.5\TeV (observed) and 2.6\TeV (expected) for an excited \PQb quark. Additionally the \Bstar production via contact interactions has been considered. Several processes contribute to the \Bstar production via CI~\cite{Baur:1989kv}, the dominant one being $\qqbar\to\PAQb\Bstar (\text{or}\,\PQb\PAQb^{\ast}) \to\PAQb\PQb\Pg$ where three jets are present in the final state.
In order to display the theory cross section for this new process on the same exclusion limit plot, only its resonant component has been considered. In particular events are selected when the reconstructed dijet originates from the \Bstar decay, and does not include the other \Pqb quark in the event, which is the case for 50 (80)\% of the events within the acceptance at 2 (4)\TeV. Including both processes in the total signal cross section (single \Bstar production and the resonant component of the \Bstar production via CI) gives a significantly more stringent 95\% \CL mass limit of 4\TeV on excited \PQb quarks.

\section{Summary}\label{sec:conclusions}

A search for heavy resonances decaying into \PQb quarks has been presented and no excess has been found over the standard model (SM) expectations. The data were collected by the CMS experiment at $\sqrt{s}=13\TeV$ during 2016--2018 and correspond to an integrated luminosity of 138\fbinv. Model-independent upper limits are set on the product of the cross section of the resonance and its branching fraction to \PQb quarks.
Signals of \PZpr bosons decaying to pairs of \PQb quarks are considered, for both the previously explored sequential standard model (SSM), and also for a new heavy vector triplet (HVT) model. The decays of \PZpr bosons in both the SSM and the HVT Model~A are excluded at 95\% confidence level for masses from 1.8 to 2.4\TeV, and limits are set on the coupling strengths of the HVT boson to SM bosons and fermions.
Signals of an excited \PQb quark are considered for a previously explored channel, $\PQb\Pg\to\Bstar\to\PQb\Pg$, and a production mode via contact interactions. The excited \PQb quark is excluded at 95\% confidence level for masses from 1.8 to 4.0\TeV. This is the most stringent exclusion of excited \PQb quarks to date.

\begin{acknowledgments}
 We congratulate our colleagues in the CERN accelerator departments for the excellent performance of the LHC and thank the technical and administrative staffs at CERN and at other CMS institutes for their contributions to the success of the CMS effort. In addition, we gratefully acknowledge the computing centers and personnel of the Worldwide LHC Computing Grid and other centers for delivering so effectively the computing infrastructure essential to our analyses. Finally, we acknowledge the enduring support for the construction and operation of the LHC, the CMS detector, and the supporting computing infrastructure provided by the following funding agencies: BMBWF and FWF (Austria); FNRS and FWO (Belgium); CNPq, CAPES, FAPERJ, FAPERGS, and FAPESP (Brazil); MES and BNSF (Bulgaria); CERN; CAS, MoST, and NSFC (China); MINCIENCIAS (Colombia); MSES and CSF (Croatia); RIF (Cyprus); SENESCYT (Ecuador); MoER, ERC PUT and ERDF (Estonia); Academy of Finland, MEC, and HIP (Finland); CEA and CNRS/IN2P3 (France); BMBF, DFG, and HGF (Germany); GSRI (Greece); NKFIH (Hungary); DAE and DST (India); IPM (Iran); SFI (Ireland); INFN (Italy); MSIP and NRF (Republic of Korea); MES (Latvia); LAS (Lithuania); MOE and UM (Malaysia); BUAP, CINVESTAV, CONACYT, LNS, SEP, and UASLP-FAI (Mexico); MOS (Montenegro); MBIE (New Zealand); PAEC (Pakistan); MES and NSC (Poland); FCT (Portugal); MESTD (Serbia); MCIN/AEI and PCTI (Spain); MOSTR (Sri Lanka); Swiss Funding Agencies (Switzerland); MST (Taipei); MHESI and NSTDA (Thailand); TUBITAK and TENMAK (Turkey); NASU (Ukraine); STFC (United Kingdom); DOE and NSF (USA).

\hyphenation{Rachada-pisek} Individuals have received support from the Marie-Curie program and the European Research Council and Horizon 2020 Grant, contract Nos.\ 675440, 724704, 752730, 758316, 765710, 824093, 884104, and COST Action CA16108 (European Union); the Leventis Foundation; the Alfred P.\ Sloan Foundation; the Alexander von Humboldt Foundation; the Belgian Federal Science Policy Office; the Fonds pour la Formation \`a la Recherche dans l'Industrie et dans l'Agriculture (FRIA-Belgium); the Agentschap voor Innovatie door Wetenschap en Technologie (IWT-Belgium); the F.R.S.-FNRS and FWO (Belgium) under the ``Excellence of Science -- EOS" -- be.h project n.\ 30820817; the Beijing Municipal Science \& Technology Commission, No. Z191100007219010; the Ministry of Education, Youth and Sports (MEYS) of the Czech Republic; the Hellenic Foundation for Research and Innovation (HFRI), Project Number 2288 (Greece); the Deutsche Forschungsgemeinschaft (DFG), under Germany's Excellence Strategy -- EXC 2121 ``Quantum Universe" -- 390833306, and under project number 400140256 - GRK2497; the Hungarian Academy of Sciences, the New National Excellence Program - \'UNKP, the NKFIH research grants K 124845, K 124850, K 128713, K 128786, K 129058, K 131991, K 133046, K 138136, K 143460, K 143477, 2020-2.2.1-ED-2021-00181, and TKP2021-NKTA-64 (Hungary); the Council of Science and Industrial Research, India; the Latvian Council of Science; the Ministry of Education and Science, project no. 2022/WK/14, and the National Science Center, contracts Opus 2021/41/B/ST2/01369 and 2021/43/B/ST2/01552 (Poland); the Funda\c{c}\~ao para a Ci\^encia e a Tecnologia, grant CEECIND/01334/2018 (Portugal); the National Priorities Research Program by Qatar National Research Fund;  MCIN/AEI/10.13039/501100011033, ERDF ``a way of making Europe", and the Programa Estatal de Fomento de la Investigaci{\'o}n Cient{\'i}fica y T{\'e}cnica de Excelencia Mar\'{\i}a de Maeztu, grant MDM-2017-0765 and Programa Severo Ochoa del Principado de Asturias (Spain); the Chulalongkorn Academic into Its 2nd Century Project Advancement Project, and the National Science, Research and Innovation Fund via the Program Management Unit for Human Resources \& Institutional Development, Research and Innovation, grant B05F650021 (Thailand); the Kavli Foundation; the Nvidia Corporation; the SuperMicro Corporation; the Welch Foundation, contract C-1845; and the Weston Havens Foundation (USA).
\end{acknowledgments}

\bibliography{auto_generated}

\cleardoublepage \appendix\section{The CMS Collaboration \label{app:collab}}\begin{sloppypar}\hyphenpenalty=5000\widowpenalty=500\clubpenalty=5000
\cmsinstitute{Yerevan Physics Institute, Yerevan, Armenia}
{\tolerance=6000
A.~Tumasyan\cmsAuthorMark{1}\cmsorcid{0009-0000-0684-6742}
\par}
\cmsinstitute{Institut f\"{u}r Hochenergiephysik, Vienna, Austria}
{\tolerance=6000
W.~Adam\cmsorcid{0000-0001-9099-4341}, J.W.~Andrejkovic, T.~Bergauer\cmsorcid{0000-0002-5786-0293}, S.~Chatterjee\cmsorcid{0000-0003-2660-0349}, K.~Damanakis\cmsorcid{0000-0001-5389-2872}, M.~Dragicevic\cmsorcid{0000-0003-1967-6783}, A.~Escalante~Del~Valle\cmsorcid{0000-0002-9702-6359}, R.~Fr\"{u}hwirth\cmsAuthorMark{2}\cmsorcid{0000-0002-0054-3369}, M.~Jeitler\cmsAuthorMark{2}\cmsorcid{0000-0002-5141-9560}, N.~Krammer\cmsorcid{0000-0002-0548-0985}, L.~Lechner\cmsorcid{0000-0002-3065-1141}, D.~Liko\cmsorcid{0000-0002-3380-473X}, I.~Mikulec\cmsorcid{0000-0003-0385-2746}, P.~Paulitsch, F.M.~Pitters, J.~Schieck\cmsAuthorMark{2}\cmsorcid{0000-0002-1058-8093}, R.~Sch\"{o}fbeck\cmsorcid{0000-0002-2332-8784}, D.~Schwarz\cmsorcid{0000-0002-3821-7331}, S.~Templ\cmsorcid{0000-0003-3137-5692}, W.~Waltenberger\cmsorcid{0000-0002-6215-7228}, C.-E.~Wulz\cmsAuthorMark{2}\cmsorcid{0000-0001-9226-5812}
% ...
\par}
\vskip\cmsinstskip
\dag:~Deceased\\
$^{1}$Also at Yerevan State University, Yerevan, Armenia\\
$^{2}$Also at TU Wien, Vienna, Austria\\
% ...
\end{sloppypar}
\end{document}